# Multifocal laser direct writing through spatial light modulation guided by scalable vector graphics

Linhan Duan,[1,#] Yueqiang Zhu,[1,#] Haoxin Bai,[1,#] Chen Zhang,[1,*] Kaige Wang,[1] Jintao Bai[1] and Wei Zhao[1,*]

[1] State Key Laboratory of Photon-Technology in Western China Energy, International Collaborative Center on Photoelectric Technology and Nano Functional Materials, Institute of Photonics & Photon Technology, Northwest University, Xi'an 710127, China;

\* Correspondence: nwuzchen@nwu.edu.cn; zwbayern@nwu.edu.cn

\# The authors contribute equally to this investigation

**Abstract:** Multifocal laser direct writing (LDW) based on phase-only spatial light modulator (SLM) can realize flexible and parallel nanofabrication with high throughput potential. In this investigation, a novel approach of combining two-photon absorption, SLM and vector path guided by scalable vector graphics (SVG) has been developed and tested preliminarily, for fast, flexible and parallel nanofabrication. Three laser focuses are independently controlled with different paths, which are according to SVG, to optimize fabrication and promote time efficiency. The minimum structure width can be as low as 74 nm. Accompanied with a translation stage, a carp structure of 18.16 μm × 24.35 μm has been fabricated. This method shows the possibility of developing LDW techniques towards full-electrical system, and provides a potential way to efficiently engrave complex structures on nanoscales.

## 1. Introduction

Two-photon lithography [1,2] is the most widely used technique in the field of micro/nano fabrication, e.g. semiconductor industry [3], micro-nanofluidics [4-7], metamaterials [8], optical information storage [9] and biomedical engineering [10-13] etc. The photosensitive molecules in the material absorb photons through two-photon absorption and initiate subsequent photopolymerization chemical reactions, which is referred as two-photon polymerization. It is a unique microfabrication technique that utilizes the nonlinear relationship between polymerization rate and radiant light intensity to produce true three-dimensional structures with feature sizes below the diffraction limit [14]. Single-point two-photon lithography was advanced initially to realize the fabrication of 3D structures [15]. The method is relatively simple and can be achieved by tuning either galvanometer or translation stages [16]. However, the single-point two-photon lithography techniques are normally time-consuming with low fabrication efficiency.

To overcome these shortcomings, a series of methods using multifocal spots to realize parallel fabrication have been developed in last decade. Multifocal spots can be generated in a variety of ways, including fixed and variable elements. Fixed elements can only produce multifocal spots with fixed number, positions and sizes [17-19]. The representative approaches are through microlens array [20] and laser interferometry [21,22], etc. However, these methods can only achieve high-throughput fabrication of the identical structures by moving the galvanometer and translation stage.

Compared to fixed elements, variable elements can produce flexible and controllable multifocal spots [23-25]. Among them, the beam shaping method based on SLM, which can modulate the phase of incident light, can generated multiple, diverse and flexibly distributed beams. It provides a high efficiency and flexible approach for laser fabrication. There are various methods for generating multifocal spots based on SLMs [26-28], such as GS algorithm [29], GSW algorithm [30], Adaptive Additive Algorithm (AA) [31] and strip segmentation phase (SSP) [32] method etc. The GS algorithm, GSW algorithm and AA are iterative algorithms. They normally take a long time in the generation of phase maps. Besides, the modulated beams generated by applying the phase maps through these iterative algorithms are normally nonuniform with undesired ripples. In contrast, SSP method which is a non-iterative method, can generate high quality phase maps for modulating multifocal spots with uniform and diverse beams. It provides an effective approach to realize fast modulation of beam for SLM-based two-photon laser direct writing.

In LDW techniques, the arrangement of path is important to improve fabrication efficiency and quality. There are some commonly used path arrangement methods in LDW, such as progressive scanning [33], zigzag scanning [34] and layer-by-layer scanning [35] for 3D. For instance, Vizsnyiczai et al [36] developed a multi-focus fabrication method realized by SLM. The phase maps were designed by GSW algorithm. The paths of the laser focuses were arranged based on the coordinate of the target. First, a multidodecahedron was virtually built up using OpenGL software and converted to voxel coordinates. Then, the voxels are manually assigned to each laser focus "ergodically". Wu et al. [37] proposed a parallel fabrication method with multi-focus array obtained via superposing opposite-ordered Bessel beams. The focal pattern and focal spot position could be rearranged by changing the order of the holograms. However, these methods have the same issues that the path should be arranged and optimized manually.

Simultaneously high flexibility and time efficiency of two-photon lithography with high resolution is what everyone has been pursuing. In this investigation, we propose a novel two-photon laser direct writing method. By combining SLM to control multi-focus movement and vector graphics to arrange path, we hope to realize independently multi-focus and flexible high-throughput nanoscopic fabrication. We hope this preliminary investigation can provide a new direction of developing SLM-based two-photon laser direct writing.

## 2. Method

The method is schematically Diagrammed in Figure 1. In the method, the fabrication of large complex structures is carried out in parallel by multifocal spots, generated and moved through SLM. The path of each laser spot is arranged according to the vector graphic of the structure that fabricated.

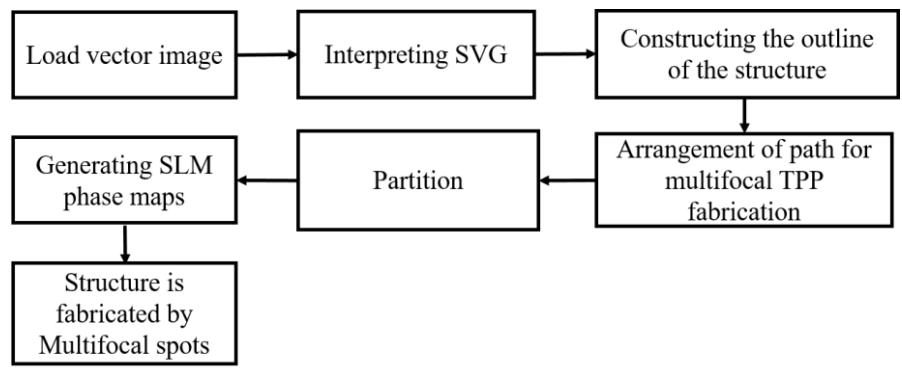

**Figure 1.** Schematic diagram of fabrication of large complex vector structures.

## 2.1 Multi-focus generated by a SLM

As is shown in Figure 2, the electric field distribution at an arbitrary point in the focal region of a high NA aberration-free lens can be calculated by Debye diffraction theory [38-41], as

$$\boldsymbol{E}(x,y,z) = \int_0^\theta \int_0^{2\pi} P(\theta)\,\boldsymbol{E}_t(\theta,\varphi) \times \exp\left\{-ik\sqrt{x^2+y^2}\sin\theta\cos[\tan^{-1}(y/x)-\varphi]\right\}$$

$$\times \exp(ikz\cos\theta)\sin\theta\,d\theta\,d\varphi$$

$$= \iint \frac{P(\theta)\boldsymbol{E}_t(\theta,\varphi)}{\cos\theta} e^{ik_z z} e^{i2\pi(\xi x+\eta y)}\,d\xi\,d\eta$$

$$= F^{-1}[P(\theta)\boldsymbol{E}_t(\theta,\varphi)e^{ik_z z}/\cos\theta\,] \tag{1}$$

where $P(\theta)$ is the pupil function of the objective, $\boldsymbol{E}_t(\theta,\varphi)$ is the transmission electric field, $\varphi$ is the azimuthal angle of the objective, $\theta = \arcsin(\frac{rNA}{Rn_t})$ is the convergence angle of the objective, $NA$ is the numerical aperture of the objective, $R$ is the maximum radius of the pupil plane behind the objective, $\theta_m$ is the maximum $\theta$, $n_t$ is the refractive index of the objective, $\xi = \cos\varphi\sin\theta/\lambda$ and $\eta = \sin\varphi\sin\theta/\lambda$ denote the spatial frequency in $x$ and $y$ directions. $\lambda$ is the wavelength in vacuum. $k_z$ is the wave vector in $z$ direction. Here, we assume the optical system follows the Abbe sinusoidal condition. The coordinates $r$ is the polar coordinate in the pupil aperture plane. $x$, $y$ and $z$ are the Cartesian coordinates of the focal region.

According to Eq. (1), the electric field after modulation is

$$E(x,y,z) = F^{-1}\left[e^{i\psi}P(\theta)\boldsymbol{E}_t(\theta,\varphi)e^{ik_z z}/\cos\theta\,\right] \tag{2}$$

where $\psi$ is the phase distribution function, which can be expressed as

$$\psi(x_0, y_0) = \frac{2\pi}{\lambda}\frac{NA}{Rn_t}(x_0\Delta x + y_0\Delta y) \tag{3}$$

where $\lambda$ is the laser wavelength, $x_0$ and $y_0$ are the coordinates of the pupil aperture plane of the objective. $\Delta x$ and $\Delta y$ are the relative displacement components relative to the original focus of the objective in the $x$ and $y$ directions of the focal plane.

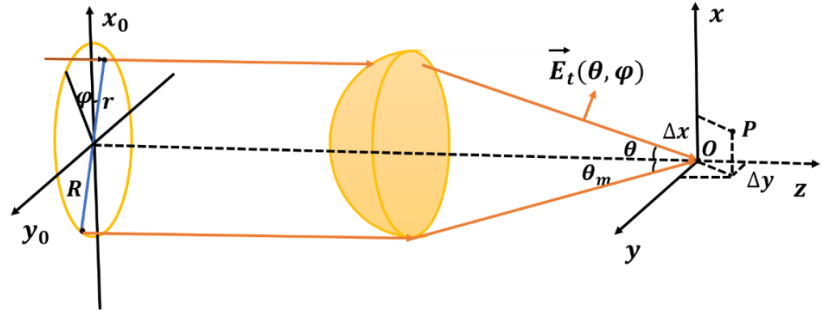

**Figure 2.** Schematic of the focused beam after passing through the objective. Here, $\varphi$ is the azimuthal angle of the objective, $r$ is polar coordinate, $\theta$ is convergence angle of the objective, $R$ is the maximum radius of the pupil plane, $\theta_m$ is the maximum convergence angle of the objective.

## 2.2 Extracting the path from SVG

A detailed description of how to extract the vector path from the SVG is Diagrammed in Figure 3. Here, a carp structure has been used as an example. Many basic components, e.g. ellipse, curves and circles etc, can be found in this SVG. When applying SVG to organize the paths, five steps are required, including (1) interpreting SVG, (2) constructing the outline of the structure, (3) arrangement of path for multifocal two-photon lithography, (4) partition, and (5) generating SLM phase maps.

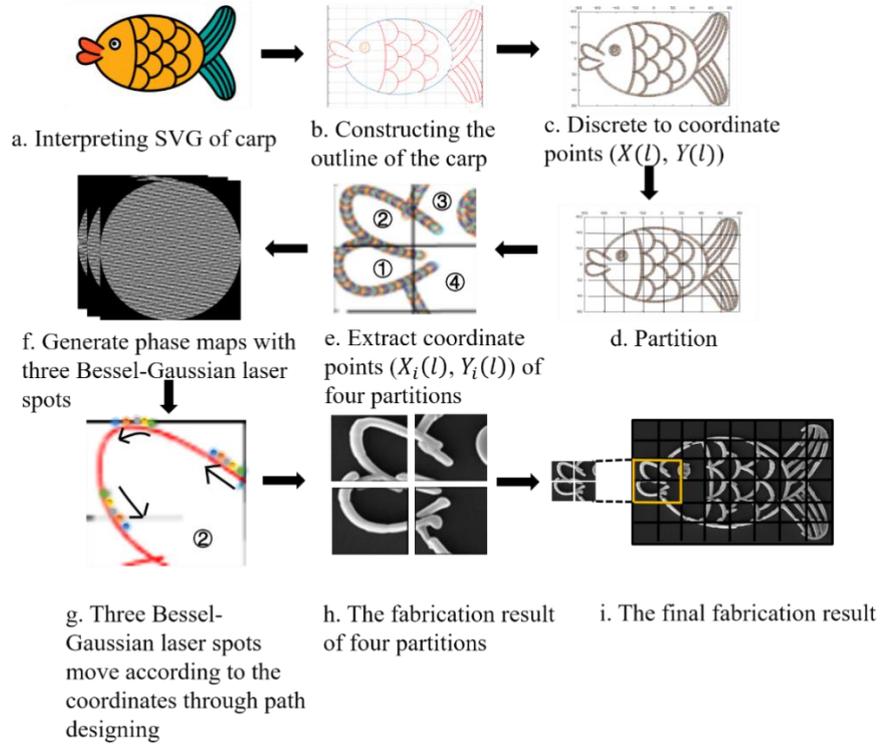

**Figure 3.** SVG decoding and fabrication flow chart with a carp structure as an example.

2.2.1. Interpreting SVG

SVG [42-44] is a language for describing two-dimensional vector graphics. Vector graphics operate on primitives such as lines, points, curves and polygons. Unlike the traditional bitmap [45] and Tiff [46] which only contain light intensity distribution information, SVG provides path information additionally. It can also be zoomed in or out without distortion and loss of quality.

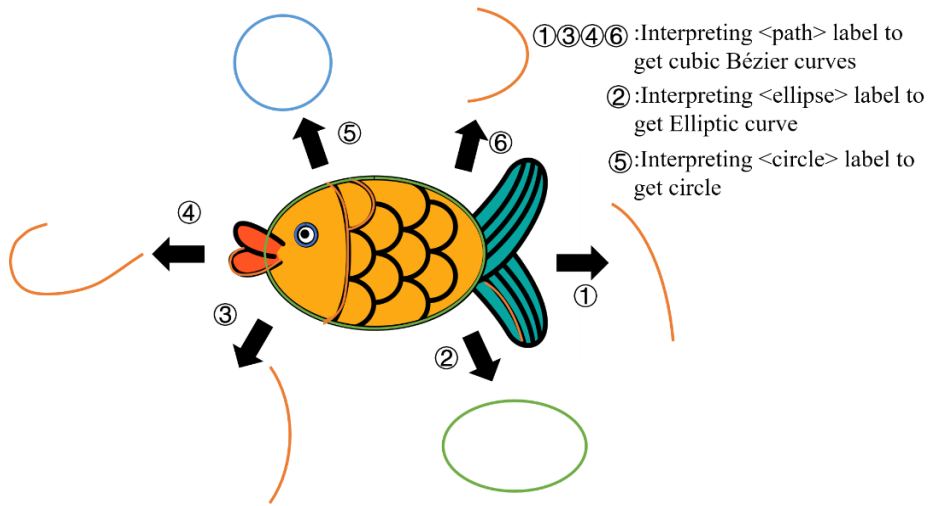

**Figure 4.** Constitution of a carp SVG as an example.

Each SVG has a corresponding text file to interpret the graph. In the text file, the labels can be interpreted as different mathematical equations to represent the curves by computer.

One typical vector graphic is plotted in Figure 4. Generally, the carp structure is constituted of head, body, scales and tail. The elliptic body is expressed by the <ellipse> label which is represented by an elliptic function as

$$\frac{(x-x_c)^2}{a^2} + \frac{(y-y_c)^2}{b^2} = 1 \qquad (4)$$

where $(x_c, y_c)$ is the coordinate of the center point of the ellipse, $a$ and $b$ are the long axis and short axis of the ellipse respectively.

Head, scales and tail are expressed by <path> labels with cubic Bézier curves, which are represented by cubic Bézier curves function $B(t) = (x, y)$, where

$$B(t) = B_0(1-t)^3 + 3B_1 t(1-t)^2 + 3B_2 t^2(1-t) + B_3 t^3, \ t \in [0,1] \qquad (5)$$

where $P_0$ is the coordinate of start point in the cubic Bézier curve, $P_1$, $P_2$ are the coordinates of two control points, respectively. $P_3$ is the coordinate of termination point in the cubic Bézier curve. $P_i = (x_i, y_i)$ with $i = 0, 1, 2, 3$ are 2D coordinates. For instance, a code <path class="cls-2" d="M58.49,10.79c7.41,0,13.43,5.3,13.43,11.83" transform="translate(0 8.81)"/> represents a cubic Bézier curve, with the coordinates $P_0 = (58.49, 10.79)$, $P_1 = (7.41, 0)$, $P_2 = (13.43, 5.3)$ and $P_3 = (13.43, 11.83)$ respectively.

The labels and the corresponding structures are summarized in Table 1. By interpreting the labels, the outline of the shape to be fabricated can be determined in the following.

2.2.2. Outline of the structure

We still use the carp structure as an example. Figure 3 (a) shows the original SVG carp pattern. Figure 3 (b) shows the outline of the carp simulated after interpreting the labels. The mouth, scales, and tail in the carp are plotted by cubic Bézier curve. The eyes and the fish body outlines in the carp are plotted by circle and ellipse, respectively. It can be seen, the outline of the carp is consistent with the SVG.

2.2.3. Arrangement of path for multifocal two-photon lithography

The outline must be transferred to discrete fabrication points for the further processing in SLM. This process is crucial on determining the fabrication efficiency and quality. It is determined by a series of parameters, including the size and resolution of the fabrication. If resolution of the fabrication is high, it is difficult to achieve a large fabrication area by SLM without moving the translation stage. A partition fabrication is required, as will elucidated in next subsection.

For high-resolution fabrication, the full width at half maxima (FWHM) of the multifocal spots must be small, with small intervals ($\Delta d$) between the fabrication points as well. According to Nyquist-Shannon Sampling Theorem [47] $\Delta d \leq d/2$, where $d = 0.61\lambda/NA$ is the FWHM of the laser spot, with $\lambda$ and $NA$ being 800 nm and 1.25 respectively. The larger $\Delta d$ is, the faster the fabrication. However, the quality may be worse. In this investigation, $\Delta d \leq d/2$ to compromise fabrication efficiency and quality. The actual value of $\Delta d$ is no larger than 0.195 μm.

Relying on the vector feature of SVG, sequential arrays of $X(l)$ (horizontal coordinates), $Y(l)$ (vertical coordinates), and their corresponding relationship arrays $P(l)$ which indicates the index of the next fabrication point, have been established. Here $l$ represents the $l$th fabrication point, $1 \leq l \leq N_{coor}$. $N_{coor}$ is the total number of the fabrication points. For example, if $P(3) = 8$, the 8th fabrication point $((X(8), Y(8))$ will be fabricated right after the 3rd fabrication point $((X(3), Y(3))$. The $P(l)$ array is generated according to the SVG.

When a structure from SVG is fabricated in parallel with multifocal spots. Defining $N_{spot}$ as the number of Bessel-Gaussian laser spots, then, there are $N_{spot}$ laser direct writing paths correspondingly. After the structure is discretized to $N_{coor}$ fabrication points, we assign the fabrication points of $[1, N_{group}]$ to the first laser spot, where $N_{group} = ceil(N_{coor}/N_{spot})$. Here, $ceil(\cdot)$ denotes a Ceiling function. Then, assign the the fabrication points of $[N_{group} + 1, 2N_{group}]$ to the second laser spot, and so forth.

**Table 1.** Coding information of SVG

| Label | Parameter | Function | Graphs and curves |
|---|---|---|---|
| <rect> | Width | | Rectangle |
| | Height | | |
| <circle> | Circle center coordinates $(x_c, y_c)$ | $(x - x_c)^2 + (y - y_c)^2 = r^2$ | Circle |
| | Radius $r$ | | |
| <ellipse> | Coordinates of the center of the ellipse $(x_c, y_c)$ | $\frac{(x - x_c)^2}{a^2} + \frac{(y - y_c)^2}{b^2} = 1$ | Ellipse |
| | Horizontal radius $a$ | | |
| | Vertical radius $b$ | | |
| <polygon> | Start coordinates $(x_1, y_1)$ | Line 1: $y = \frac{y_2-y_1}{x_2-x_1} \cdot (x - x_1) + y_1$ | Polygon |
| | $(x_2, y_2)\ldots (x_n, y_n)$ $(n \geq 3)$ | Line 2: $y = \frac{y_3-y_2}{x_3-x_2} \cdot (x - x_2) + y_2$ …… | |
| | End coordinates $(x_n, y_n)$ | Line $n$: $y = \frac{y_n-y_1}{x_n-x_1} \cdot (x - x_1) + y_1$ | |
| <line> | Start coordinates $(x_1, y_1)$ | $y = \frac{y_2 - y_1}{x_2 - x_1} \cdot (x - x_1) + y_1$ | Line |
| | End coordinates $(x_2, y_2)$ | | |
| <polyline> | Start coordinates $(x_1, y_1)$ | Line 1: $y = \frac{y_2-y_1}{x_2-x_1} \cdot (x - x_1) + y_1$ | Polyline |
| | $(x_2, y_2) \ldots (x_{n-1}, y_{n-1})$ $(n \geq 3)$ | Line 2: $y = \frac{y_3-y_2}{x_3-x_2} \cdot (x - x_2) + y_2$ …… | |
| | End coordinates $(x_n, y_n)$ | Line $n-1$: $y = \frac{y_n-y_{n-1}}{x_n-x_{n-1}} \cdot (x - x_{n-1}) + y_{n-1}$ | |
| <path> | M/m | $(x_0, y_0)$ | Initial coordinate of outline |
| | L/l | Line: $y = \frac{y_2-y_1}{x_2-x_1} \cdot (x - x_1) + y_1$ | Straight line |
| | V/v | Line: $x = a$ | Vertical line |
| | H/h | Line: $y = a$ | Horizontal line |
| | C/c S/s | $x = x_0(1 - t)^3 + 3x_1 t(1 - t)^2 + 3x_2 t^2(1 - t) + x_3 t^3, t \in [0,1]$ $y = y_0(1 - t)^3 + 3y_1 t(1 - t)^2 + 3y_2 t^2(1 - t) + y_3 t^3, t \in [0,1]$ | Cubic Bezier curve |
| | Q/q T/t | $x = x_0(1 - t)^2 + 2x_1 t(1 - t) + x_2 t^2, t \in [0,1]$ $y = y_0(1 - t)^2 + 2y_1 t(1 - t) + y_2 t^2, t \in [0,1]$ | Quadratic Bezier curve |
| | A/a | $x = x_c + a\cos\theta$ $y = y_c + b\sin\theta$ $\theta \in (\theta_1, \theta_2)$ | Elliptical Arc |
| | Z/z | $(x_n, y_n)$ | End coordinate of outline |

2.2.4. Partition

In SLM-based laser direct writing, the multifocal spots can be moved by SLM only without moving the translation stage. However, the movement range of the multifocal spots is determined by the optical system. If a high fabrication resolution is required, we need to use a high numerical aperture (NA) objective lens. Accordingly, the range of multifocal spots that can be moved by SLM is quite limited. For instance, to fabricate structures with O(100 nm) resolution, an objective lens of $NA = 1.25$ is applicable. Restricted by the small diameter of the objective lens aperture, the movement range of multifocal spots is only $\pm 1.5$ μm, relative to the center of the modulation field. Thus, if a large area target is to be fabricated, the original fabrication points must be divided into partitions before generating SLM phase maps, as shown in Figure 3(d).

The number ($N_{part}$), size and position of the partitions should be determined according to the performance of optical system and the area of target. During partition, all the $N_{coor}$ fabrication points are assigned to each partition (Figure 3(e)), re-assigned as $X_i(l)$, $Y_i(l)$ and $P_i(l)$ which are

$$X_i(l) = X(l) - X_{C,i}$$
$$Y_i(l) = Y(l) - Y_{C,i}$$
(7)

where $1 \le l \le N_{coor,i}$, $i$ denotes the ith partition, $X_{C,i}$ and $Y_{C,i}$ represent the center coordinates of the $i$th partition. If $N_{spot}$ multifocal spots are modulated in the parallel fabrication, the $N_{coor,i}$ fabrication points will be assigned to the $N_{spot}$ as average as possible. Equation (3) in this case then reads.

$$\psi(x_0, y_0) = \frac{2\pi}{\lambda} \frac{NA}{Rn_t} (x_0 X_i(l) + y_0 Y_i(l))$$
(8)

It should be noted that, $P_i(l)$ needed to be re-assigned according to $P_i(1)$ which is normally the fabrication point on boundary.

In the fabrication of the carp structure which is 18.16 μm by 24.35 μm, $N_{spot} = 3$ Bessel-Gaussian laser spots have been modulated simultaneously. With an $NA = 1.25$ objective lens, each partition are 3 μm × 3 μm. Therefore, the carp were divided into $6 \times 8$ partitions, as shown in Figure 3(d).

2.2.5. Generating SLM phase maps

(A) For the targets which has a smaller size than the movement range of the multifocal spots, no partition is needed. The phase maps of multifocal spots shown in Figure 3(f) can be generated directly by SLM according to Equation (3), by

$$\Delta x = X(l)$$
$$\Delta y = Y(l)$$
(9)

Overall, a number of $N_{pm} = ceil(N_{coor}/N_{spot})$ phase maps are generated, with $N_{spot}$ fabrication points or less in each phase map. Here, $ceil(\cdot)$ denotes a Ceiling function. After loading all the phase maps in sequences into the SLM, the multifocal spots move according to the coordinates, as Diagramed in Figure 3(g). Then, with properly control the exposure time of individual phase map, the target can be parallelly fabricated by two-photon lithograph.

(B) For the targets which has a larger size than the movement range of the multifocal spots, $N_{part}$ partitions are needed. The phase maps are generated according to the index of partition first, i.e. the phase maps in the first partition is generated in sequence. Then, the phase maps in the second partition are generated immediately following the last phase map of the first partition. By repeating the progress above, all the phase maps can be generated. The generation sequence is as follows.

$$M_1(1), M_1(2), \cdots, M_1(N_{pm,1}), M_2(1), M_2(2), \cdots, M_2(N_{pm,2}),$$
$$\cdots, M_{N_{part}}(1), M_{N_{part}}(2), \cdots, M_{N_{part}}(N_{pm,N_{part}})$$
(10)

In each partition, the phase maps can be generated according to Equation (8) and follow the steps in case (A). By synchronizing the movement of translation stage, loading of phase maps, and laser exposure time, the structures in each partition can be fabricated in sequence, as shown in Figure 3 (h). Finally, the carp structure with large area is fabricated, as shown in Figure 3 (i).

## 3. Experimental setup

The LDW system is Diagramed in in Figure 5. The light source is a femtosecond laser (Coherent, Chameleon Ultra II) with a center wavelength of 800 nm. The power of the incident laser is first adjusted by a Glan prism (GP) and a half-wave plate (HWP) (Thorlabs, 700 nm-1000nm), then, filtered by a spatial light filter (SLF) and collimated by two lenses (L1 and L2) to improve the beam quality. A mechanical shutter (MS)(SH05, Thorlabs) has been applied to accurately control the exposure time of the light source, the MS response speed is about 4.08 ms. The laser subsequently passes through the second HWP (Thorlabs, 650 nm-1000nm), the polarization direction of beam to be consistent with the long axis of the liquid crystal on silicon (LCoS) SLM (PLUTO-2-VIS-016, Holoeye). The expanded laser is loaded with the corresponding phase information after passing through LCoS SLM. Then the modulated laser is reflected through a mirror to the dichroic mirror (DM) (Chroma, ZT532rdc_NIR), after reflected on the dichroic mirror, the laser is focused into photoresist by a high-NA objective lens (OL) (ACHN100XOP 100x NA 1.25, Olympus) for LDW. The fabrication process is monitored through a CMOS camera (MER-2000-19U3C-L, Daheng Optoelectronics) in real time.

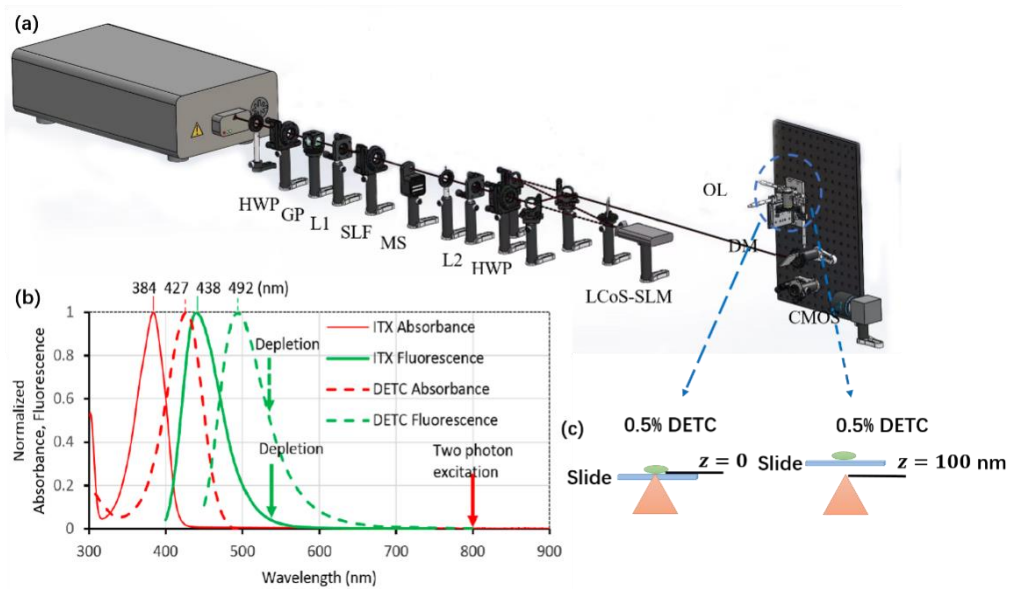

**Figure 5.** (a) Diagram of the experiment setup. HWP half-wave plate, GP glan prism, SLF spatial light filter, lens L1 and L2 composed beam expander, MS mechanical shutter, LCoS liquid crystal plate on silicon SLM, DM dichroic mirror, OL objective lens, R reflector, CMOS camera. (b) Absorbance and fluorescence spectra of 0.5 wt% 7-diethylamino3-thenoylcoumarin (DETC) [48]. (c) The position of the modulated laser in the photoresist. When $z = 0$, the modulated laser is located at the junction of the photoresist and the slide. When $z = 100$ nm, the sample stage is 100 nm up through the three-dimensional nano translation table compared to $z = 0$.

In this investigation, the photoresist is a mixture of photoinitiator DETC and pentaerythritol triacrylate (PETA) monomer, which are mixed in a mass ratio of 1:199. After stirred through a magnetic stirrer for 12 hours, a clean and transparent yellow-green viscous liquid, i.e. 0.5% DETC is prepared. As shown in Figure 5 (a), the photoresist has a wide absorption band around 427 nm and emits fluorescence around 492 nm.

## 4. Result

For the targets which has a smaller size than the movement range of the multifocal spots, no partition is needed, while for the targets which has a larger size than the movement range of the multifocal spots, partition are needed. Here, we show how the fabrication is carried out and the preliminary results.

*4.1 Fabrication results of concentric circle and star structures in small area*

In the first case, a concentric circle structure is fabricated by a single Bessel-Gaussian laser spot, to show the feasibility of beam control by SLM. The FWHM of the laser spot is $d = 0.39$ μm and the light field modulation range is 3 μm × 3 μm. The interval between the fabrication points $\Delta d = 0.07$ μm. The power is 3 mW and the exposure time for a single phase map is 17 ms. The fabrication process is illustrated in Figure 6(a-f). As shown in Figure 6(g1-g3), a concentric circle structure is successfully fabricated by the method. The difference, the method is ultrasensitive to the vertical position of the laser spot. Two fabrication results on two different planes, e.g. $z = 0$ (Figure 6(g1)), $z = 100$ nm (Figure 6(g2)), have been shown. The vertical positions of the planes have been Diagramed in Figure 5(c). By slightly moving the laser spot in vertical direction, the fabrication width is decreased from 95 nm to 74 nm. One problem is that the width of the inner circle of the concentric circle is significantly larger than that of the outer circle, according to overexposure. This could be attributed to the different modulation efficiency of the modulated laser spot at different fabrication points.

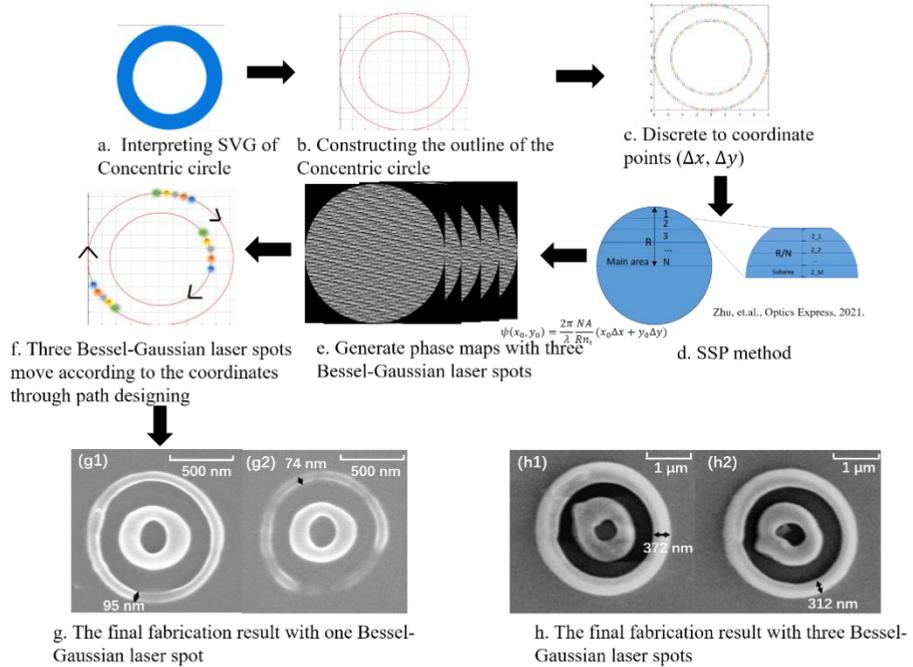

**Figure 6.** Fabrication of a small area concentric circle structure with one Bessel-Gaussian laser spot and three Bessel-Gaussian laser spots, respectively. The fabrication process is illustrated in (a-f).and (g - h) show the fabrication results. Here, $d = 0.39$ μm and $\Delta d = 0.07$ μm. The exposure time for a single phase map is 17 ms (g) Only a single laser spot is used, where the laser power is 3 mW. (g1) $z = 0$, and (g2) $z = 100$ nm. (h) Three laser spots are used simultaneously, where the laser power is 12 mW. (h1) $z = 0$ and (h2) $z = 100$ nm.

Subsequently, the concentric circle structure is fabricated simultaneously by three Bessel-Gaussian laser spots. Here, we only want to validate the feasibility of the method, thus, more laser spots have not been applied. In this case, $d = 0.39$ μm and $\Delta d = 0.07$ µm. The light field modulation range is still 3 µm × 3 µm. The laser power is 12 mW and the exposure time for a single phase map is also 17 ms. The fabrication results are shown in Figure 6(h1) and (h2). Although the concentric circle structures have been fabricated, the width of structure is reduced apparently relative to that by a single laser spot. The minimum width is 312 nm. In the fabrication of inner circles, the structures are nonuniform with worse fabrication quality, relative to that by a single laser spot. However, the uniformity of the outer circles in Figure 6(h) are better than that of Figure 6(g). From both Figure 6(g) and (h), it can be seen, above a distance from the fabrication center, uniform and high-resolution fabrication with either single of multifocal spots can be achieved simultaneously.

A relatively more complex structure—star which is described by cubic Bézier curve, is also fabricated by three Bessel-Gaussian laser spots. The FWHM of the laser spot is $d = 0.39$ µm and the light field modulation range is 3 µm × 3 µm. The interval between the fabrication points $\Delta d = 0.07$ µm. The power is 10 mW and the exposure time for a single phase map is 17 ms. Here, three star structures of different sizes are fabricated, as shown in Figure 7(g). From Figure 7(g1)-(g3), as the size of the fabricated star structure is increased from 2.28 µm to 2.84 µm, the fabrication resolution increases gradually, with a decreasing width to 248 nm. It can be seen that at the same laser power, if the size of the star structure is larger, the width of the star structure is smaller, i.e. a high fabrication resolution.

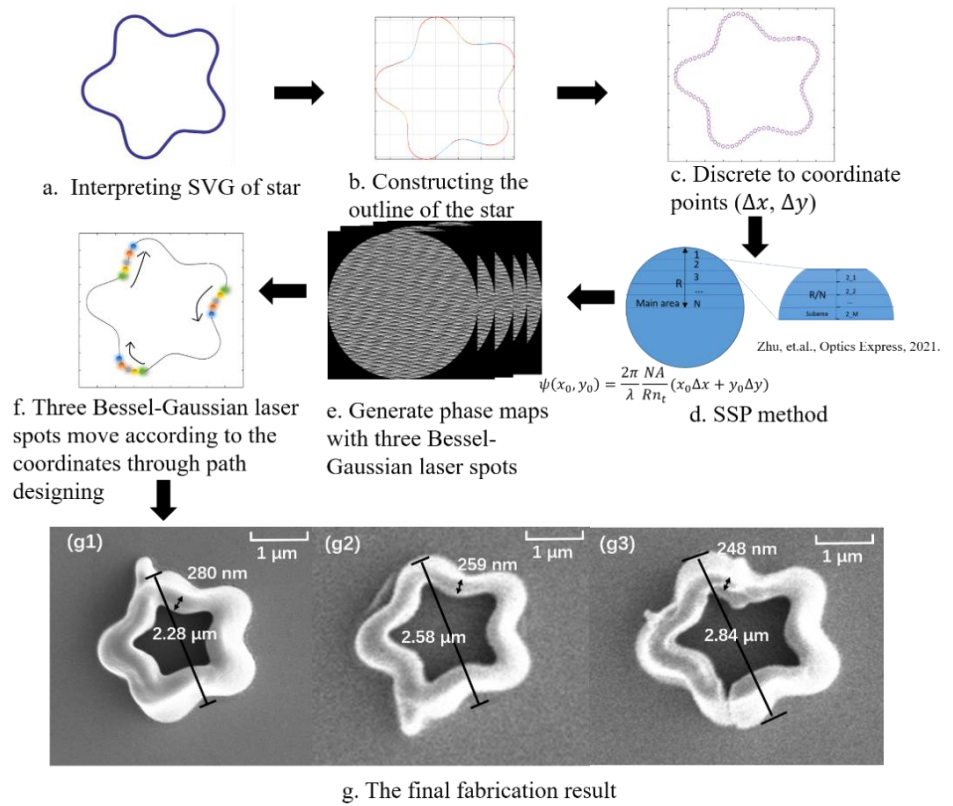

**Figure 7.** Fabrication of a small area star structure with three Bessel-Gaussian laser spots. The fabrication process is illustrated in (a-f).and (g) show the fabrication results. Here, $d = 0.39$ µm, $\Delta d = 0.07$ µm. The laser power is 10 mW and the exposure time for a single phase map is 17 ms. The sizes of the structures in (g1)-(g3) are 2.28 µm, 2.58 µm and 2.84 µm respectively.

*4.2 Fabrication of carp structure in large area*

In this section, a fabrication of carp structure is introduced. It has an area of 18.16 μm × 24.35 μm, which is larger than the modulation area of 3 μm × 3 μm, therefore, partition is needed in this case. Here, we still use three Bessel-Gaussian laser spots, with $d = 0.39$ μm and $\Delta d = 0.07$ μm. The exposure time for a single phase map is 17 ms. As shown in Figure 8, under a laser power of 8.5 mW, the typical width of the carp structure is 483 nm. The complete structure has been fabricated, even though the structures at the boundaries of the partitions are not smooth, as shown in Figure 3(h). The different modulation efficiencies of the modulated laser spots at different positions could be one of the reasons. Besides, in the fabrication process, due to the jitter of the 3D nano translation stage, the two adjacent fabrication partitions are slightly misaligned. The accuracy of the connection between adjacent curves needs to be strengthened.

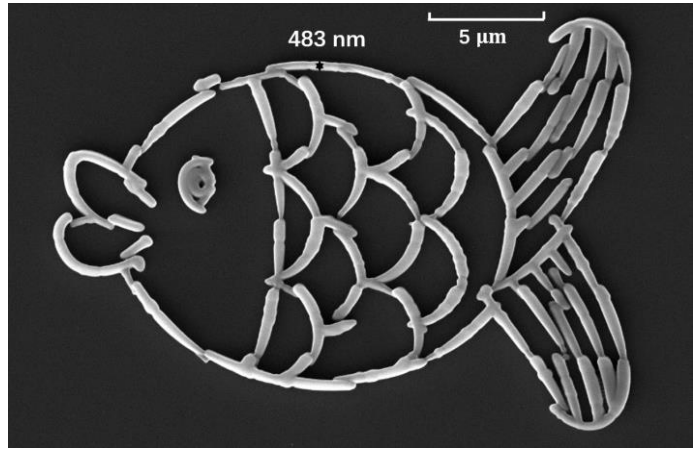

**Figure 8.** Fabrication of a large area carp structure with three Bessel-Gaussian laser spots. The size of the carp structure is 18.16 μm × 24.35 μm. Here, $d = 0.39$ μm, $\Delta d = 0.07$ μm. The laser power is 8.5 mW and the exposure time for a single phase map is 17 ms.

## 5. Discussion

Although the approach advanced in this investigation has been proved to be effective, from the fabrication results, there are still numerous problems remained. On the one hand, the fabrication results are influenced by the modulation efficiency ($\delta = P_m/P_0$) of the modulated laser spot at different positions, where $P_0$ and $P_m$ are the laser powers before and after modulation. This issue significantly affects the power of modulated laser spot and the fabricated structure. Figure 9 shows the modulation efficiency of laser spot at different positions. Generally, $\delta$ is slightly higher in the central part of the modulation range, while lower near the edges. The minimum $\delta$ is 0.78, while the maximum $\delta$ is 1, corresponding to the unmodulated laser spot. On the other hand, the fixed $\Delta d$ can lead to a stronger overlapping in the central part. Both of the issues lead to a wider fabrication structures in the central part than that near the edges. As an example in Figure 6(g) and (h), we can see a larger width and stronger non-uniformity in the inner ring of the concentric circle than that of the outer ring. Similarly, in Figure 7(g), a larger star structure has a smaller structure width.

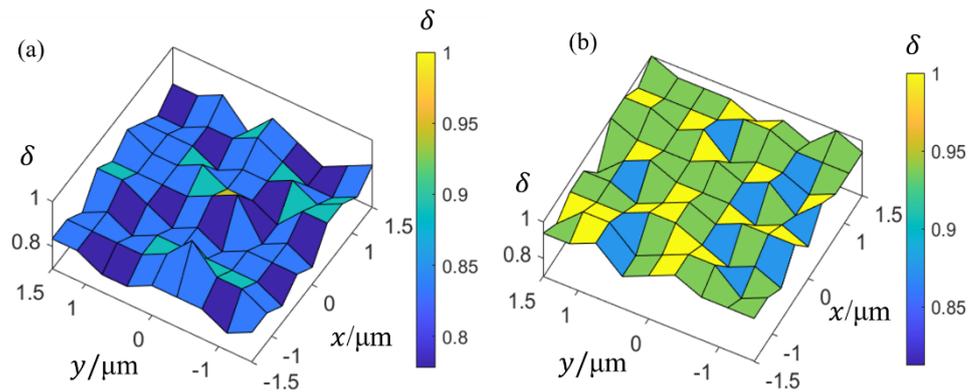

**Figure 9.** Distribution of modulation efficiency at different input laser powers. (a) 3 mW and (b) 4 mW.

### 6. Conclusion

In this investigation, we developed a novel LDW method by combining two-photon absorption, SLM and vector path guided by SVG for fast, flexible and parallel nanofabrication. As a preliminary test of the method, three laser focuses are independently controlled with three different paths to optimize fabrication and promote time efficiency. In the fabrication with a single laser spot, a minimum width of 74 nm has been reached. While for the fabrication with three laser spots, accompanied with a translation stage, a carp structure of 18.16 μm × 24.35 μm has been fabricated with a typical structure width of 483 nm.

Although the current results are still defective, indicating the method is still far from being developed, the method shows potency to change the inherent lithography method, make it possible to efficiently fabricate complex structures. The current method exhibits a possibility of developing LDW techniques towards full-electrical system and realizes the lithography technology of engrave complex structures quickly and efficiently.

**Author Contributions:**

**Funding:** This research was funded by National Major Scientific Research Instrument Development Project of China (Grant No. 51927804); National Natural Science Foundation of China (Grant No., 61405159); Science and Technology Innovation Team Project of Shaanxi Province (No. S2018-ZC-TD-0061).

**Conflicts of Interest:** The authors declare no conflict of interest.